\documentstyle[prb,aps,psfig]{revtex} 
\begin{document}

\draft

\title{$s$ and $d$-wave symmetries of the solutions of the Eliashberg equations}

\author{
G. Santi$^{(a)}$\cite{address}, 
T. Jarlborg$^{(a)}$, 
M. Peter$^{(a)}$, 
M. Weger$^{(b)}$}
\address{
(a) DPMC, Universit\'{e} de Gen\`eve,\\
CH-1211 Gen\`eve 4, Switzerland\\
(b) Racah Institute of Physics, Hebrew University,\\
Jerusalem, Israel}

\date{Received \today}

\maketitle

\begin{abstract}

We examine the different possible symmetries of the superconducting gap obtained by solving the Eliashberg equations. 
We consider an electron-phonon interaction in a strong coupling scenario. 
The Coulomb pseudopotential plays the crucial role of providing the repulsion needed to favour the $d$-wave symmetry. 
But the key parameter that allows very anisotropic solutions even with very strong coupling is the small angular range of the interaction due to predominantly electron-phonon forward scattering that is found in the high-$T_c$ superconductors. 
We find both $s$ and $d$-wave solutions whose stability depends mainly on the angular range of the interaction.

\end{abstract}

\vspace{0.5cm}



\section{Introduction} 

In the last few years, much experimental and theoretical effort has been spent in trying to understand the mechanism responsible for superconductivity in the high-$T_c$ superconductors. 
Such a mechanism might be the BCS electron-phonon interaction, the exchange of paramagnons, or some Coulomb interaction processes as in the $t$-$J$ model or in the Pines picture \cite{pines}. 
The 
symmetry of the gap has been given much attention since it may clarify some
aspects of the superconducting mechanism. 
It seems now quite established, at least in YBCO, that the symmetry of the gap is $d$-wave \cite{squid}. 

Our aim in this paper is to show that a standard electron-phonon attractive interaction in conjunction with a weak Coulomb repulsion can lead to a $d$-wave symmetry of the superconducting gap $\Delta$. 
This goes against the common
idea that an electron-phonon mechanism can only lead to an $s$-wave symmetry. 
Indeed, if the sign of the interaction does not change, there is a theorem in linear algebra saying that the gap will not have nodes \cite{pokrovski}. 
Furthermore, it has been shown \cite{combescot} that a large electron-phonon coupling would wash out any anisotropy in the gap. 
It is true that in normal metals, the electron-phonon interaction is local due to the strong screening, thereby allowing averaging over the Fermi surface (FS) \cite{peter}. 
In the cuprates, the situation is different. 
The dielectric constant of the ionic background is very large for small frequencies, i.e. 
$\epsilon_{ion}(\omega \rightarrow 0) \approx 40 - 60$ in YBCO \cite{humlicek} (what we denote by $\epsilon_{ion}$ is the dielectric constant obtained after substracting the Drude contribution of the conduction electrons at frequencies lower than the frequencies of the transverse optical modes; in this frequency range, $\epsilon$ is nearly independent of $\omega$; this quantity is sometimes denoted $\epsilon_\infty$ \cite{reagor}). 
Since the response of the ions to an external charge is so strong, the response of the electrons, given normally by $\epsilon_{el}(q) = 1 + (\kappa_{TF}/q)^2$ with $\kappa_{TF}^2 = 4 \pi e^2 N(\epsilon_F)$, will be decreased. 
Therefore, the dressed electronic dielectric constant $\epsilon^{dressed}_{el}$ can be written as $\epsilon^{dressed}_{el}(q) = 1 + (\epsilon_{el}-1)/\epsilon_{ion} = 1 + \kappa_{TF}^2/(q^2\epsilon_{ion})$. 
The range of the interaction in $k$-space for a normal metal is given by $\kappa_{TF}$ ($\approx 2k_F$), while in our case,
 $\kappa_{TF}/\sqrt{\epsilon_{ion}} \approx 0.3k_F$. 
Therefore, the electron-phonon scattering is predominantly forward, i.e. the angular range of the interaction (for electrons at the FS) $\delta\theta \approx \kappa_{TF}/(\epsilon_{ion}k_F) \approx \pi/10$ is very small. 
This electron-phonon interaction then allows for a very anisotropic $s$-wave gap and even a $d$-wave one, due to the repulsive Coulomb interaction (since the sign of the total interaction is not the same everywhere).
Apart from these considerations, it seems worthwhile to examine the electron-phonon interaction in the cuprates since there have been some experimental indications from Point Contact Spectroscopy that the phonons play a crucial role in the superconductivity of YBCO \cite{vedeneev}.

We therefore consider a strong electron-phonon interaction through our generalized Eliashberg scheme \cite{hpa,zphys,annphys}. 
In this scheme we solve the Eliashberg equations \cite{eliashberg} for an electron-phonon interaction having a very strong $\bf k$-dependence of its matrix elements $g({\bf k},{\bf k'})$ that enters in the phonon propagator $D({\bf k},{\bf k'}; \omega_n-\omega_{n'})$ through eq. (\ref{eq:phprop}). 
Apart from the angular dependence discussed above, the same arguments about the short range of the interaction in momentum space  also apply to the dependence on the bare quasiparticles energy $\xi_k$ (i.e. on the component of the momentum $\bf k$ perpendicular to the FS \cite{miami,ssp}). 
Therefore, we consider an interaction that is cut off for high energy transfers, i.e. $g({\bf k},{\bf k'}) \neq 0$ only if $|\xi_k-\xi_{k'}| \leq \xi_{cut}$, where $\xi_{cut}$ is the energy of the cutoff. 
Furthermore, we shall exclude the study of the $p$-wave pairing from our analysis since we shall only consider the case of singlet pairing. 

However, we want to emphasize that the existence of a $d$-wave solution of Eliashberg equation does not rely on a cutoff at low energy \cite{ssp} since as we shall see in the following section, even the weak coupling treatment in which the $\xi_k$-dependence is completely neglected can exhibit a $d$-wave solution.
Furthermore, we show in section IV, that the strong coupling case without
cutoff also gives $d$-wave solutions.

\section{Weak Coupling Situation}
\label{weakcoup}

Although a strong coupling calculation is needed to treat correctly the case of the High-$T_c$ Superconductors, the important parameters, as well as the main trend, can be put forward by the standard weak coupling theory. 
We will consider here an isotropic FS, and the electron-phonon interaction as being responsible for superconductivity. 
We want to study the effect of the anisotropy of the {\em interaction}, as well as of the temperature on the symmetry of the gap in the superconducting state.

We solved numerically the standard gap equation \cite{pokrovski} restricted in 2 dimensions

\begin{equation}
\label{eq:bcsgap}
\Delta(\theta) = \frac{1}{2\pi} \int_{-\pi}^{\pi} d\theta' 
\int_{-\omega_{max}}^{\omega_{max}} d\xi_{k'}
\frac{\Delta(\theta')}{\sqrt{\xi^2_{k'} + \Delta^2(\theta')}} \tanh 
\left( \frac{\sqrt{\xi^2_{k'} + \Delta^2(\theta')}}{2T} \right)
[V(\theta,\theta') - \mu^*(\theta,\theta')]
\end{equation}

where $\omega_{max}$ is the maximum phonon frequency. 
The Coulomb pseudopotential $\mu^*$ is chosen as constant for simplicity (in this section). 
We consider a very anisotropic electron-phonon interaction $V(\theta,\theta')$ corresponding to a predominantly forward scattering situation. 
We chose to modelize the behaviour of $V(\theta,\theta')$, where the angles $\theta$ and
$\theta'$ refer to points $k$ and $k'$ on the FS (cf fig 1), with a gaussian :

\begin{equation}
\label{eq:gauss}
V(\theta,\theta') = \frac{\lambda}{2 \delta\theta \, \mbox{erfc} (\frac{\pi}{\delta\theta})} \exp 
\left( -\frac{(\theta-\theta')^2}{\delta\theta^2} \right),
\end{equation}

the parameter $\delta\theta$ determining the angular range of the interaction.

For $T=0$ and a small gap $\Delta(\theta) < \omega_{max}$ (i.e. a small electron-phonon coupling constant $\lambda$), eq. (\ref{eq:bcsgap}) can be written

\begin{equation}
\label{eq:bcsgap2}
\Delta(\theta) = \frac{1}{2\pi} \int_{-\pi}^{\pi} d\theta' 
\Delta(\theta') \ln \left| \frac{2\omega_{max}}{\Delta(\theta')} \right| 
[V(\theta,\theta') - \mu^*(\theta,\theta')] .
\end{equation}

\begin{figure}[h]
\centerline{
\psfig{figure=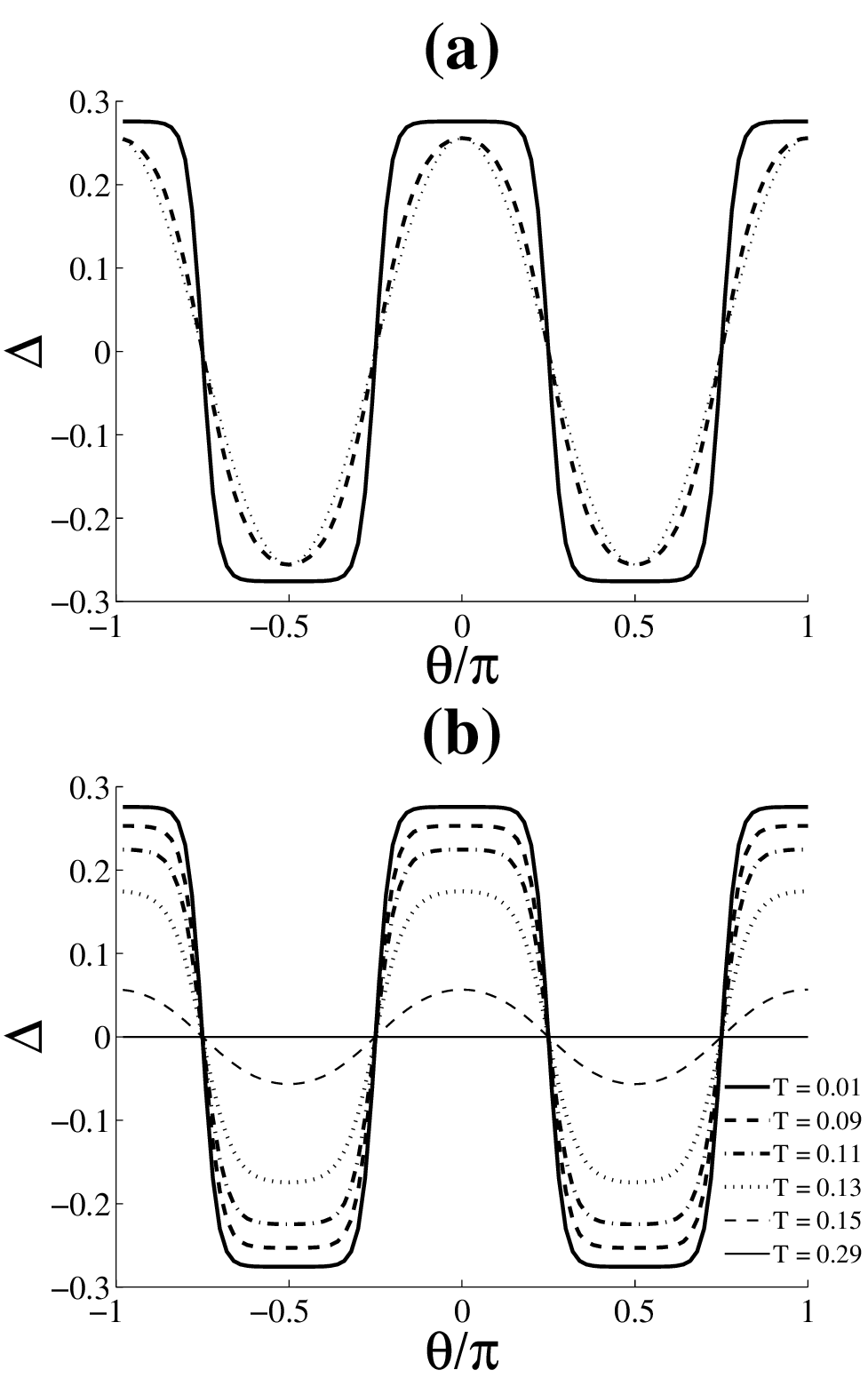,height=19cm}
}
\vspace{1cm}
\caption{\label{f:gapshape} Shape of the gap $\Delta_d(\theta)$ with the coupling parameters $\lambda=0.5$ and $\mu^*=0.02$. 
{\bf (a)} For 2 different angular ranges $\delta\theta = \pi/30$ (plain line), $\delta\theta = \pi/10$ (dashed line) and a cosine function is also plotted for comparison (pointed line); and {\bf (b)} for different temperatures, i.e. $T=0.01, 0.09, 0.11, 0.13, 0.15$ and 0.29 (with $\delta\theta=\pi/30$).}
\end{figure}

\begin{figure}[h]
\centerline{
\psfig{figure=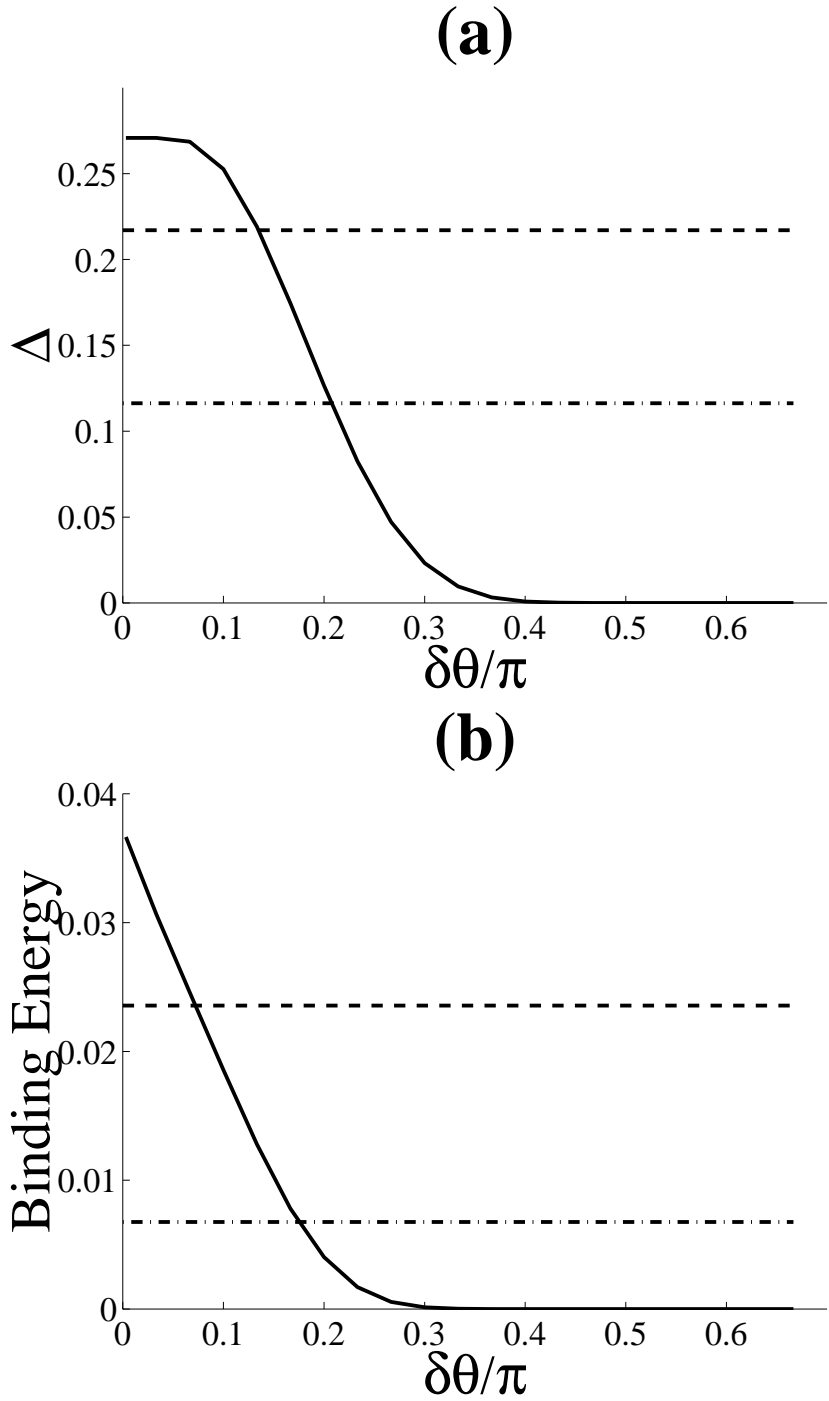,height=19cm}
}
\vspace{1cm}
\caption{\label{f:gapvsdth} Dependence of the solution on the angular range $\delta\theta$. 
{\bf (a)} Maximum of the $d$-wave gap $\Delta_d$ vs $\delta\theta$ and values of the $s$-wave gap $\Delta_s$ for $\mu^*=0.05$ (dashed line) and $\mu^*=0.15$ (dash-dot line). 
{\bf (b)} Binding energy (with $N(\epsilon_F)=1$) for the same cases as before.}
\end{figure}

\begin{figure}[h]
\centerline{
\psfig{figure=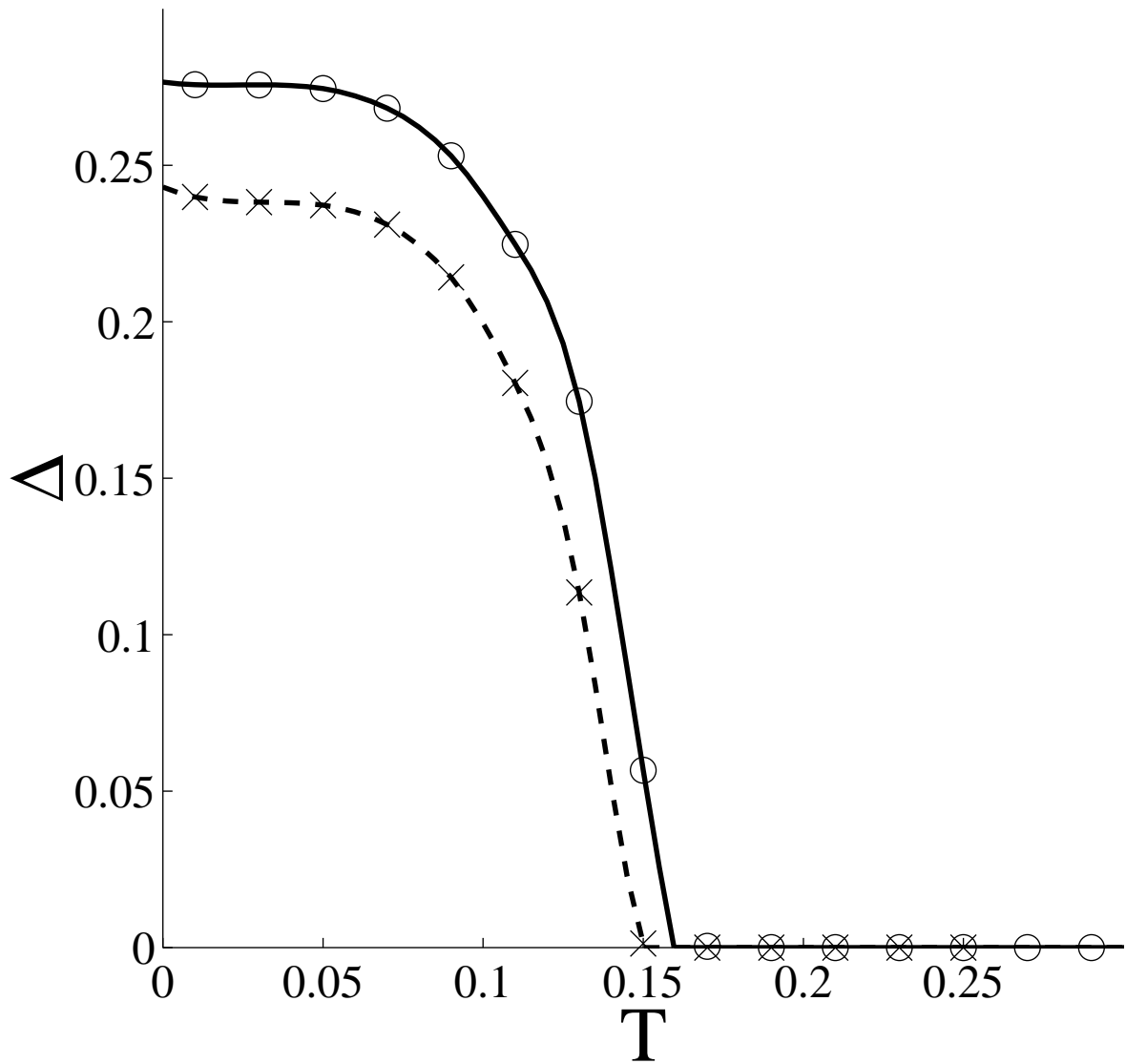,height=15cm}
}
\vspace{1cm}
\caption{\label{f:gapvst} Temperature-dependence of the maximum of the gap for the $d$-wave case with $\delta\theta=\pi/30$ (plain line and circles) and for the $s$-wave with $\mu^*=0.02$ (dashed line and crosses).}
\end{figure}

We find that there always exists an $s$-wave solution to eqs. (\ref{eq:bcsgap}) or (\ref{eq:bcsgap2}), but nevertheless a $d$-wave solution can also be found provided the angular range of the interaction $\delta\theta$ is small, i.e. the electron-phonon scattering is predominantly forward. 
We show the shape of the gap $\Delta_d(\theta)$ for the $d$-wave solution for 2 different values of $\delta\theta$ in fig. \ref{f:gapshape}a. 
We see that the shape of $\Delta_d(\theta)$ becomes square-like when $\delta\theta \rightarrow 0$, whereas it is cos-like for larger $\delta\theta$. 
The temperature also has the same effect of softening the shape of the gap (see fig. \ref{f:gapshape}b), diminishing thereby the anisotropy. 
Fig. \ref{f:gapvsdth}a shows how $\Delta_d \equiv \max \Delta_d(\theta)$ depends on the angular range $\delta\theta$. 
We see that $\Delta_d$ decreases rapidly with $\delta\theta$, falling to zero when $\delta\theta \approx \pi/2$. 
The values of $\Delta_s$ for 2 different values of $\mu^*$ are also shown for comparison. 
We plotted the corresponding binding energy $E_{bind} = \frac{N(\epsilon_F)}{2} \frac{1}{2\pi} \int_{-\pi}^{\pi}{d\theta \Delta^2(\theta)}$ in fig. \ref{f:gapvsdth}b (we set $N(\epsilon_F)=1$ for simplicity). 
These 2 figures show clearly that a large $\delta\theta$ will favour $s$-wave and a small one the $d$-wave. 
The crossover $\delta\theta_c$ between these two symmetries is given by $E_{bind}^{(s)}(\delta\theta_c) = E_{bind}^{(d)}(\delta\theta_c)$. 
When $\delta\theta > \delta\theta_c$, the system will prefer $s$-wave even if the $s$-gap is smaller then the maximum value of the $d$-wave gap. 
Then, in the case presented here, the crossover, for $\mu^* = 0.05$, will take place at $\delta\theta_c \approx \pi/14$, but the maximum $d$-wave gap will remain larger than the $s$-wave gap until $\delta\theta \approx \pi/8$. 
As it can be seen from fig. \ref{f:gapvst}, the temperature dependence of the gap is quite BCS-like. 
We observed that $T_c$ and the shape of $\Delta(T)$ do not depend significantly neither on the angular range $\delta\theta$, nor on the symmetry of the gap (this was done by varying $T$, keeping $\delta\theta$ constant). 
Therefore a  temperature-induced crossover from $d$ to $s$-wave, with constant (temperature-independent) parameters is unlikely. 
However, an increase of the temperature might broaden the range of the interaction $\delta\theta$ since the predominantly forward scattering may be somehow washed out by thermal effects. 
Then, since a relatively small change in $\delta\theta$ can have a very big effect on the magnitude of the gap (as it is seen from fig. \ref{f:gapvsdth}), the temperature-effects will be seen through the dependence of $\delta\theta$ on $T$. 
Therefore one has the possibility of a crossover from $d$ to $s$-wave induced by the temperature. 
When $\delta\theta$ is very small it is clear that other symmetries higher than d can be possible, but here we have not studied
such solutions.

We also tried another possibility for the attractive coupling $V(\theta,\theta')$. 
We considered that the strength of the on-site coupling could vary with the angle $\theta$, in order to favour the apparition of nodes in the gap. 
We then chose to replace $\lambda$ in eq. (\ref{eq:gauss}) by $\lambda \varphi(\theta) \varphi(\theta')$, with $\varphi(\theta) = \cos^{2n} 2\theta$ or $\varphi(\theta) = \frac{1}{2}(1 + \cos^{2n} 2\theta)$. 
Indeed the strength of both of these coupling would be smaller at the place on the FS where a node of the $d$-wave solution should be found. 
The solutions for different sets of parameters showed that although the shape of the gap $\Delta(\theta)$ was altered significantly, the main qualitative results about the existence of a $d$-wave solution remained unchanged, i.e. the important parameter allowing for a $d$-wave solution is $\delta\theta$. 
Indeed if $\delta\theta$ is too large (\raisebox{-1ex}{$\stackrel{\textstyle >}{\sim}$} $\pi/10$), not only the $d$-wave solution disappears, but the superconductivity as well, i.e. $\Delta(\theta) = 0$.

From this weak coupling calculation, we clearly see that the key physical property that allows the $d$-wave symmetry is the tendency to forward scattering. 
Indeed, since there are only few electrons side-scattered, the gap at a point of the FS will only have very little influence on the gap in the neighbouring region. 
Then it is possible to have a very anisotropic solution.

\section{Strong Coupling Case}

The value of $2\Delta/k_B T_c$ in the cuprate clearly shows that these materials are in the strong coupling regime. 
We have therefore solved the Eliashberg equations \cite{eliashberg} to find the superconducting state. 
Since the FS of the cuprates is mainly 2 dimensional, we consider the form of the Eliashberg equations that is already integrated along the remaining momentum direction, i.e. these equations specialized to 2 dimensional momentum space. 
Instead of using $(k_x,k_y)$ as momentum variables, we will use the more convenient ``polar-like'' $\mbox{\boldmath $\kappa$} \equiv (\theta,\xi_k)$ where $\theta$ is the angle and $\xi_k \equiv \epsilon_k-\epsilon_F$. 
As we already did in previous publications \cite{hpa,zphys,annphys,miami}, we do not integrate out the dependence on the momentum $\xi_k$. 
The Eliashberg equations can then be written in the Matsubara frequencies formalism ($\omega_{n}=\pi T(2n+1)$):

\begin{eqnarray}
\Phi(\mbox{\boldmath $\kappa$},\omega_{n})&=& T\sum_{n'}
\int\!\!\!\!\int d^2\mbox{\boldmath $\kappa'$} \frac{N(\mbox{\boldmath $\kappa'$})(D(\mbox{\boldmath $\kappa$},\mbox{\boldmath $\kappa'$},\omega_{n}-\omega_{n'})-\mu(\mbox{\boldmath $\kappa$},\mbox{\boldmath $\kappa'$}))
\: \Phi(\mbox{\boldmath $\kappa'$},\omega_{n'})}{\Omega}
\label{eq:eliash1} \\
(X(\mbox{\boldmath $\kappa$},\omega_{n})-1) \: \xi_k & = & - T \sum_{n'}
\int\!\!\!\!\int d^2\mbox{\boldmath $\kappa'$} \frac{N(\mbox{\boldmath $\kappa'$}) (D(\mbox{\boldmath $\kappa$},\mbox{\boldmath $\kappa'$},\omega_{n}-\omega_{n'})-\mu(\mbox{\boldmath $\kappa$},\mbox{\boldmath $\kappa'$}))
\: X(\mbox{\boldmath $\kappa'$},\omega_{n'}) \: \xi_{k'}}{\Omega}
\label{eq:eliash2} \\ 
(Z(\mbox{\boldmath $\kappa$},\omega_{n})-1) \: \omega_{n} & = &  T\sum_{n'}
\int\!\!\!\!\int d^2\mbox{\boldmath $\kappa'$} \frac{N(\mbox{\boldmath $\kappa'$})(D(\mbox{\boldmath $\kappa$},\mbox{\boldmath $\kappa'$},\omega_{n}-\omega_{n'})-\mu(\mbox{\boldmath $\kappa$},\mbox{\boldmath $\kappa'$}))
\: Z(\mbox{\boldmath $\kappa'$},\omega_{n'}) \: \omega_{n'}}{\Omega}
\label{eq:eliash3} \\
\Omega&=&(Z(\mbox{\boldmath $\kappa'$},\omega_{n'}) \omega_{n'})^2 + 
(X(\mbox{\boldmath $\kappa'$},\omega_{n'}) \xi_{k'})^2 + \Phi(\mbox{\boldmath $\kappa'$},\omega_{n'})^2
\label{eq:eliash4}
\end{eqnarray}

where $D(\mbox{\boldmath $\kappa$},\mbox{\boldmath $\kappa'$},\omega_{n}-\omega_{n'})$ is the phononic propagator given by (we assume here an Einstein spectrum for simplicity)

\begin{equation}
\label{eq:phprop}
D(\mbox{\boldmath $\kappa$},\mbox{\boldmath $\kappa'$},\omega_{n}-\omega_{n'})=g(\mbox{\boldmath $\kappa$},\mbox{\boldmath $\kappa'$})^2
\frac{2\omega_{ph}}{(\omega_{n}-\omega_{n'})^2+\omega_{ph}^2},
\end{equation}

and $\mu(\mbox{\boldmath $\kappa$},\mbox{\boldmath $\kappa'$})$ is the Coulomb repulsive pseudopotential that we will consider as mainly energy-independent for simplicity (we think it should depend on $\xi_k$, $\xi_{k'}$ \cite{zba}; although this effect is not negligible, we will neglect it here for simplicity).
Therefore, we have chosen $\mu(\mbox{\boldmath $\kappa$},\mbox{\boldmath $\kappa'$}) = \mu(\theta,\theta')$. 
In practice we use weighted Matsubara
frequencies when solving eq. (\ref{eq:eliash1}-\ref{eq:eliash4}) in order to speed up the computations \cite{jarlborg}.

As we already pointed out \cite{annphys}, it is possible to explain
some typical high $T_c$ properties from a model where the electron-phonon interaction is severely cut off along the $\xi_k$ direction. 
Indeed, the dielectric constant of the ionic background $\epsilon_{ion}$ is anomalously big at small energies \cite{humlicek}, i.e. $\epsilon_{ion}(\xi \rightarrow 0) \approx 40 - 60$ whereas it is normal at higher energies, that is of the order of 2. 
This will have the effect of reducing the electronic screening at the low energies, thereby enhancing the electron-phonon interaction there \cite{polonica}. 
We modelize this behaviour through a cutoff for low energies in the electron-phonon coupling constant:

\begin{equation}
g(\mbox{\boldmath $\kappa$},\mbox{\boldmath $\kappa'$}) = 
g(\theta,\theta') \; \Theta(|\xi_k-\xi_{k'}| - \xi_{cut}(\theta,\theta'))
\end{equation}

where $\Theta(x)$ is the Heaviside step function and $\xi_{cut}(\theta,\theta')$ determines the cutoff and is typically of order $\omega_{ph}/2$ (except where the elastic scattering rate by defects and impurities $\tau_{el}^{-1}$ is big, forcing the cutoff to be larger (see below)).

For the angular dependence of the interaction, guided by the above weak coupling calculation, we choose the interaction $g(\theta,\theta')$ to be rather strongly peaked around the angle $\theta$, which means that $\delta\theta$ in eq. (\ref{eq:gauss}) is small. 
Therefore $\lambda(\theta,\theta') = 2 N(\epsilon_F) g^2((\theta,\theta') / \omega_{ph}$ corresponds to the situation of forward scattering that favours anisotropic solutions of the gap.

In our formulation of the Eliashberg equations, $Z(\mbox{\boldmath $\kappa$},\omega_{n})$ is the standard mass renormalization and $X(\mbox{\boldmath $\kappa$},\omega_{n})$ is the renormalization of the bands due to the interaction effects. 
Because of the small cutoff, $X$ depends strongly on the temperature and the gap. 
The gap is given by

\begin{equation}
\label{eq:truegap}
\Delta(\mbox{\boldmath $\kappa$},\omega_{n}) = 
\frac{\Phi(\mbox{\boldmath $\kappa$},\omega_{n})}{Z(\mbox{\boldmath $\kappa$},\omega_{n})}.
\end{equation}

The strong $\xi_k$-dependence of the interaction has very interesting effects on the gap parameter $\Phi$ and on the renormalization functions $X$ and $Z$ that we studied in some length elsewhere \cite{hpa,zphys,annphys}. 
In the context of the study of the symmetries of the gap, these effects are not very important because they are somewhat hidden by the effects of the anisotropy of the coupling. 
Nevertheless, we wanted to keep the same formalism in order to continue to explore the richness of the solutions it can bring, and in order
to be consistent with our earlier approach \cite{hpa,zphys,annphys}.

\section{Results}

\begin{figure}[h]
\centerline{
\psfig{figure=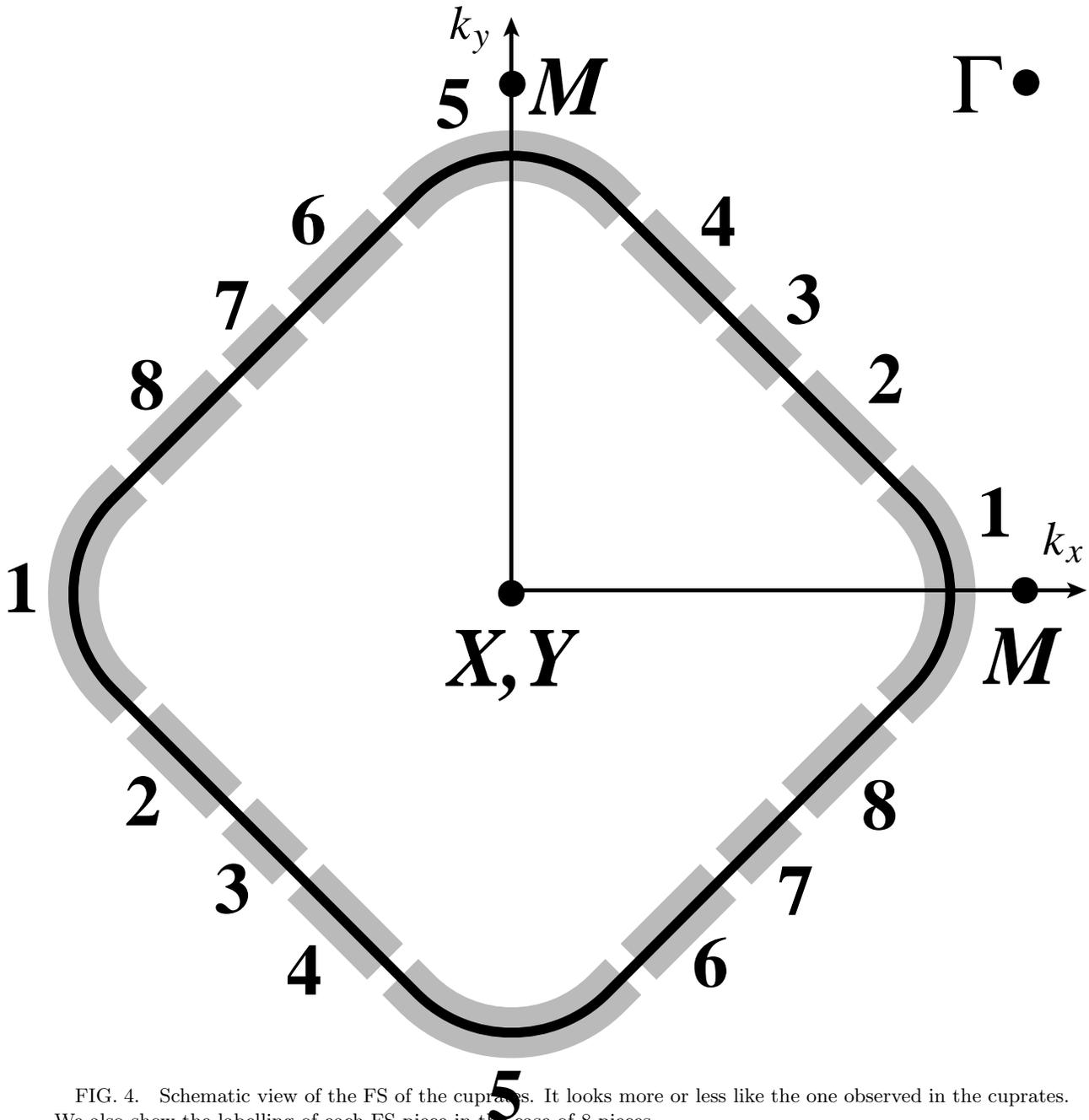,height=18cm}
}
\caption{\label{f:fs} Schematic view of the FS of the cuprates. 
It looks more or less like the one observed in the cuprates. 
We also show the labelling of each FS piece in the case of 8 pieces.}
\end{figure}

We solved the set of equations (\ref{eq:eliash1}-\ref{eq:eliash4}) numerically by iterations. 
For the discretization of the integration over $\xi_{k'}$ we used a mesh with 101 points, and for the integration over $\theta'$, we split our simplified FS (see below and fig. \ref{f:fs}) in 4 or 8 pieces in an angular range from 0 to $\pi$ (one half of the total FS). 
The sum over the Matsubara frequencies $\omega_n$ was carried on a set of frequencies equivalent to the 68 first ones \cite{jarlborg} which means that the range along the imaginary axis was at least 8 times the phonon frequency $\omega_{ph}$.

We take the same shape of the FS as in earlier works \cite{annphys,miami}. 
This is the simplest case one could imagine beyond the very rough approximation of an isotropic FS. 
It looks like a square with rounded corners (see fig. \ref{f:fs}), and may bear some resemblance with the FS observed by ARPES measurements in YBCO \cite{arpes} or in BiSCCO \cite{campuzano}. 
We consider a FS consisting mainly of 2 different kinds of pieces: the planes and the corners. 
The corners are near a Van Hove singularity (thereby having a large density of states) and $v_F$ (perpendicular to the z-axis) 
is then much smaller than on the planes. 
Consequently the elastic scattering rate $\tau_{el}^{-1}$ is larger on the corners, smearing out the electronic momentum ($\xi_k$) and preventing the cutoff $\xi_{cut}$ from being very small. 
This is why we will choose a much bigger cutoff for the corners than for the planes. 
In order to be able to reproduce all usual symmetries of the gap (i.e. $s$ and $d$-wave), we need to split the FS in at least 4 pieces in a range of $\pi$ (half the FS). 
Fig. \ref{f:fs} shows the splitting of the FS in 8 pieces, the 4 pieces case being simply obtained by grouping the 3 pieces on the ``plane'' part of the FS in a single piece.

For simplicity we set $\omega_{ph}=1$ and $N(\mbox{\boldmath $\kappa$})=1 \; \; \forall \; \mbox{\boldmath $\kappa$}$ (the {\boldmath $\kappa$}-dependence of $N(\mbox{\boldmath $\kappa$})$ is implicitly incorporated in $g(\mbox{\boldmath $\kappa$},\mbox{\boldmath $\kappa'$})$). 
All the energies are given in units of $\omega_{ph}$. 
Since we only consider a small number of pieces, specified by their angular position $\theta_i, i=1,\ldots,4$ or 8, we will use a matrix notation, denoting $g^2(\theta_i,\theta_j)$ by $g^2_{ij}$. 
All these matrices $g^2_{ij}, {\xi_{cut}}_{ij}$ and $\mu_{ij}$ are obviously symmetric. 
We start with the description of the parameters of the 8 pieces case, the case with 4 pieces follows the same general philosophy and its parameters will be given subsequently. 
As was mentioned above, the cutoff $\xi_{cut}$ on the corners cannot be as small as the one on the planes. 
We have chosen ${\xi_{cut}}_{11} = {\xi_{cut}}_{55} = 2$, and ${\xi_{cut}}_{ii} = 0.4$ for all other diagonal matrix elements, i.e. for $i \neq 1$ and 5; the non-diagonal elements were set to $\xi_{ij} = 2 \;\;\; \forall i \neq j$. 
The angular dependence of the electron-phonon coupling $\lambda_{ij} = 2 g^2_{ij}$ (for our choice of $\omega_{ph}$ and $N(\mbox{\boldmath $\kappa$})$) is basically as described by eq. (\ref{eq:gauss}). 
The situation is just a little bit complicated by the anisotropy of the FS and our choice for the splitting in which the pieces have not the same weight. 
The diagonal elements were chosen as follows: 
$\lambda_{11} = \lambda_{55} = 2$ (corners), $\lambda_{22} = \lambda_{44} = \lambda_{66} = \lambda_{88} = 1.2$ (part of the planes neighbouring the corners), and $\lambda_{33} = \lambda_{77} = 0.6$ (central part of the planes). 
The non-diagonal elements are rapidly decreasing with the distance in momentum space: $\lambda_{12} = \lambda_{18} = \lambda_{45} = \lambda_{56} = 0.7$ for the coupling between neighbouring corners and planes, $\lambda_{23} = \lambda_{34} = \lambda_{67} = \lambda_{78} = 0.4$ between neighbouring parts of the planes, and $\lambda_{ij} = 0.2$ for the rest. 
This corresponds approximately to a gaussian shape in the angular interaction with $\delta\theta \approx \pi/8$. 
The Coulomb repulsion is, as we already said, mainly constant, but with only some variations taking in account the different weights of the different parts of the FS and the repulsive interaction between corners as seen in neutron experiments \cite{neutron}and described by the theory of Monthoux et al \cite{pines}. 
Therefore, we have a more repulsive interaction (that will favour the $d$-wave solution) between electrons belonging to pieces separated by an angle $\Delta\theta = |\theta-\theta'|$ of about $\pi/2$. 
In the light of this, we set the matrix elements of $\mu$ as follows: the corner-corner interaction $\mu_{c-c}(\Delta\theta=\pi/2) = \mu_{15} = 0.4$, the plane-plane one $\mu_{p-p}(\Delta\theta=\pi/2) = \mu_{26} = \mu_{37} = \mu_{48} = 0.3$, the corner-plane one (for pieces separated by $3\pi/8$) $\mu_{c-p} (\Delta\theta=3\pi/8) = \mu_{14} = \mu_{16} = \mu_{25} = 0.3$, and the rest $\mu_{ij} = 0.1$ except for the 2 diagonal elements standing for the on-site repulsion of the corners $\mu_{c-c}(\Delta\theta=0) = \mu_{11} = \mu_{55} = 0.25$ (to take in account the weight of these pieces).

The results of this calculation with 8 FS pieces are shown in fig. \ref{f:8p} for $T \rightarrow$0 case (for technical reasons the temperature was set very low, $T=0.02\omega_{ph}$). 
The gap function $\Phi(\mbox{\boldmath $\kappa$},\omega_{n})$, as well as the renormalization functions $X(\mbox{\boldmath $\kappa$},\omega_{n})$ and $Z(\mbox{\boldmath $\kappa$},\omega_{n})$ are plotted as functions of $\xi_k$ for each of the 8 FS pieces (denoted by the angle $\theta_j = \frac{\pi}{8} (j-1), j = 1,\ldots,8$) and for $n=0$, since we believe that the low $\omega$ behaviour of these functions is roughly constant \cite{annphys}, so that it is satisfactory to have their values at $\omega=0$ (which in turn is approximately the value for the $\omega_{n=0}=\pi T$ Matsubara frequency). 
Both $s$- and $d$-wave solutions were found for this set of parameters (dashed and continuous lines respectively), the symmetry of the solution depending only on the choice for the symmetry of the initial condition. 
The true thermodynamic gap of the $s$-wave solution $\Delta_s = \Phi/Z$ (eq. (\ref{eq:truegap})) is quite anisotropic, although less than its function $\Phi$, since the behaviour of the function $Z$ tends to diminish this anisotropy \cite{combescot}. 
There are 3 different values for the gap $\Delta_s$ at the FS: $\Delta_{s1} = \Delta_{s5} = 0.48$, $\Delta_{s2} = \Delta_{s4} = \Delta_{s6} = \Delta_{s8} = 0.35$ and $\Delta_{s3} = \Delta_{s7} = 0.32$ (in units of $\omega_{ph}$), where $\Delta_{sj} \equiv \Delta_s(\theta_j, \xi_k=0, \omega_{n=0})$. 
On the other hand, the maximum value of the gap for the $d$-wave solution $\Delta_d^{max}$ is much larger, i.e. $\Delta_d^{max} \equiv \max \Delta_d(\theta_j) = 0.88$ (on the corners). 
We could then expect from our weak coupling considerations that the system will prefer the $d$-wave symmetry since such a large difference between the magnitudes of the $s$- and $d$-wave gaps indicates that we are not so close to the crossover described in fig. \ref{f:gapvsdth}.

In order to show the effect of the choice of the initial condition, we studied a case with only 4 FS pieces (grouping the 3 pieces of the plane in 1) for 3 different initial conditions. 
We considered an $s$-wave starting gap $\Phi_{0,s}(\mbox{\boldmath $\kappa$},\omega_{n}) = 1 \;\; \forall \; \mbox{\boldmath $\kappa$}, n$ and 2 $d$-wave ones corresponding to a $d_{x^2-y^2}$, i.e. $\Phi_{0,d_{x^2-y^2}}(\mbox{\boldmath $\kappa$},\omega_{n}) = \cos 2\theta \;\; \forall \; \xi_k, n$ and to a $d_{xy}$, i.e. $\Phi_{0,d_{xy}}(\mbox{\boldmath $\kappa$},\omega_{n}) = \sin 2\theta \;\; \forall \; \xi_k, n$. 
The solutions for the gap parameter $\Phi(\mbox{\boldmath $\kappa$},\omega_{n})$ are shown in fig. \ref{f:4p}. 
The coupling parameters are set with the same philosophy as for the 8 FS pieces case and are as follows:
\[
\lambda_{ij} = \left( \begin{array}{cccc}
   2 & 0.6 & 0.3 & 0.6 \\
   0.6 & 2 & 0.6 & 0.3 \\
   0.3 & 0.6 & 2 & 0.6 \\
   0.6 & 0.3 & 0.6 & 2 \\
\end{array} \right), \;\;\;\;\;\;\;
{\xi_{cut}}_{ij} = \left( \begin{array}{cccc}
   2 & 2 & 2 & 2 \\
   2 & 0.4 & 2 & 2 \\
   2 & 2 & 2 & 2 \\
   2 & 2 & 2 & 0.4 \\
\end{array} \right), \;\;\;\;\;\;\;
\mu_{ij} = \left( \begin{array}{cccc}
   0.25 & 0.25 & 0.4 & 0.25 \\
   0.25 & 0.25 & 0.25 & 0.4 \\
   0.4 & 0.25 & 0.25 & 0.25 \\
   0.25 & 0.4 & 0.25 & 0.25 \\
\end{array} \right),
\]

where the pieces numbered 1 and 3 stand for the corners and the two others for the planes. 
Here we chose a somewhat less anisotropic electron-phonon interaction than for the 8 FS pieces case. 
The range of the interaction can be estimated as $\delta\theta \approx \pi/3$ (but the angular dependence is not exactly gaussian).

The solution shown in fig. \ref{f:4p}a possesses an $s$-wave symmetry, as expected. 
The anisotropy of the gap is similar to the one of the calculation
with 8 pieces. 
The true gap $\Delta_s$ as defined in eq. (\ref{eq:truegap}) is (for $\xi_k=0$ and $n=0$) $\Delta_{s1} = \Delta_{s3} = 0.44$ on the corners and $\Delta_{s2} = \Delta_{s4} = 0.32$ on the planes. 
The $d_{x^2-y^2}$ solution in fig. \ref{f:4p}b also has the same symmetry as the initial gap. 
In fact, this solution has not exactly the $d_{x^2-y^2}$ symmetry since the nodes are not found at $\theta=\pi/4$ but they are probably slightly displaced 
somewhere in the seams between pieces 1 and 2, and 3 and 4 (corner and plane). 
This is due to the small number of pieces considered. 
The maximum value of the gap is smaller than for the corresponding $s$-wave solution, i.e. $\Delta_d^{max} = 0.36$, as it can be expected since the angular range of the interaction $\delta\theta$ is larger. 
The last result presented in fig. \ref{f:4p}c is very similar to the previous one.
The only visible difference with the gap parameter of fig. \ref{f:4p}b is that the sign of its component on the plane (which is small) is reversed. This seems to be the only effect of the $d_{xy}$-like initial condition. 

We have also observed asymmetrical solutions of the Eliashberg equations (neither $d$ nor $s$-like), i.e. of lower symmetry than the interaction and the FS that indicate a spontaneous symmetry breaking (a kind of Jahn-Teller effect). We will discuss the physical meaning of these solutions in a forthcoming communication.

Finally we also considered the 4 pieces without cutoff parameters in order to
see if the question of $s$- vs $d$-wave symmetry was affected by this. 
The new solutions are qualitatively the same as when the cutoff was introduced.
Clear $d$-symmetry are found, but there is no dispersion of $\Phi$ with $\xi_k$.
The plots corresponding to fig. \ref{f:4p} consist just of straight, horizontal lines, and we do not show them here.
Thus we conclude that the cutoff parameters are not crucial for the 
appearance of $d$-wave vs $s$-waves. 
Still, the cutoffs lead to other characteristics
of the gap etc.

\section{Discussion}

As we already noted in a previous work \cite{miami}, the choice of the symmetry of the initial gap used to start the iterative procedure is crucial for the symmetry of the solution. 
Indeed, there always exists an $s$-wave solution of the Eliashberg equations (\ref{eq:eliash1}-\ref{eq:eliash3}), but a $d$-wave solution can also be found (depending this time on the values of the other parameters) provided that the iterations were started with a $d$-wave like gap. 
This duality of solutions can be easily understood. 
In the limiting case where $\delta\theta \rightarrow 0$ (and $\mu \rightarrow 0$), i.e. where each FS piece is entirely decoupled from the others, the phase of the gap parameter $\Phi$ can be chosen freely on each piece. 
Then there is no constraint on the symmetry of the gap, or, in other words, the $s$-wave and $d$-wave solutions (or sometimes
even higher symmetry) are completely degenerate. 
In the case considered here, the interaction between electrons of different pieces, although not strictly zero, is quite small. 
We are then in a nearly degenerate state where both $s$ and $d$ wave solutions can coexist, but where the relative phase of the gap parameter between pieces is determined. 
In the light of this discussion, it seems rather astonishing that we were not able to get a clear $d_{xy}$-wave solution (fig. \ref{f:4p}c). 
Indeed, the matrix $\lambda_{ij}$ is symmetrical with respect to the exchange of plane and corner, and therefore the symmetry of the result should depend on the symmetry of the initial condition. 
But the asymmetry between planes and corners is in the cutoff matrix $\xi_{ij}$, reducing sufficiently the average attractive coupling on the planes to impose the $d_{x^2-y^2}$-wave solution over the $d_{xy}$ one. 
(A similar effect can be obtained without cutoff by setting smaller values for the diagonal element $\lambda_{ii}$ of the planes than for the one of the corners).

\begin{figure}[h]
\centerline{
\psfig{figure=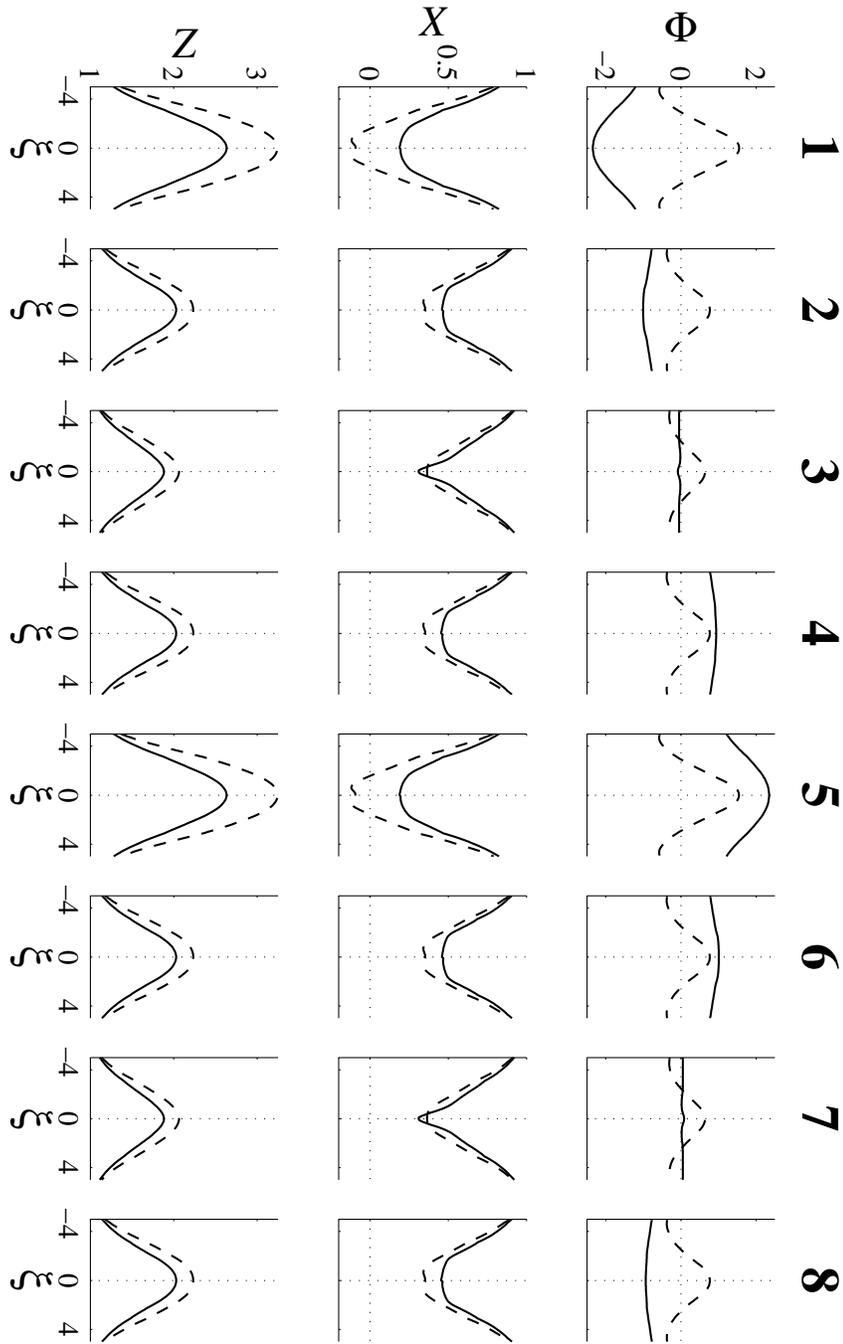,height=18cm}
}
\vspace{1cm}
\caption{\label{f:8p} Solutions of the Eliashberg equations $\Phi$, $X$ and $Z$ as functions of the bare quasiparticles energy $\xi_k$ for each of the 8 FS pieces. 
The $s$-wave symmetry (dashed line) and $d$-wave symmetry (plain line) are shown. 
The parameters describing the electron-phonon coupling and the Coulomb pseudopotential are discussed in the text.}
\end{figure}

\begin{figure}[h]
\centerline{
\psfig{figure=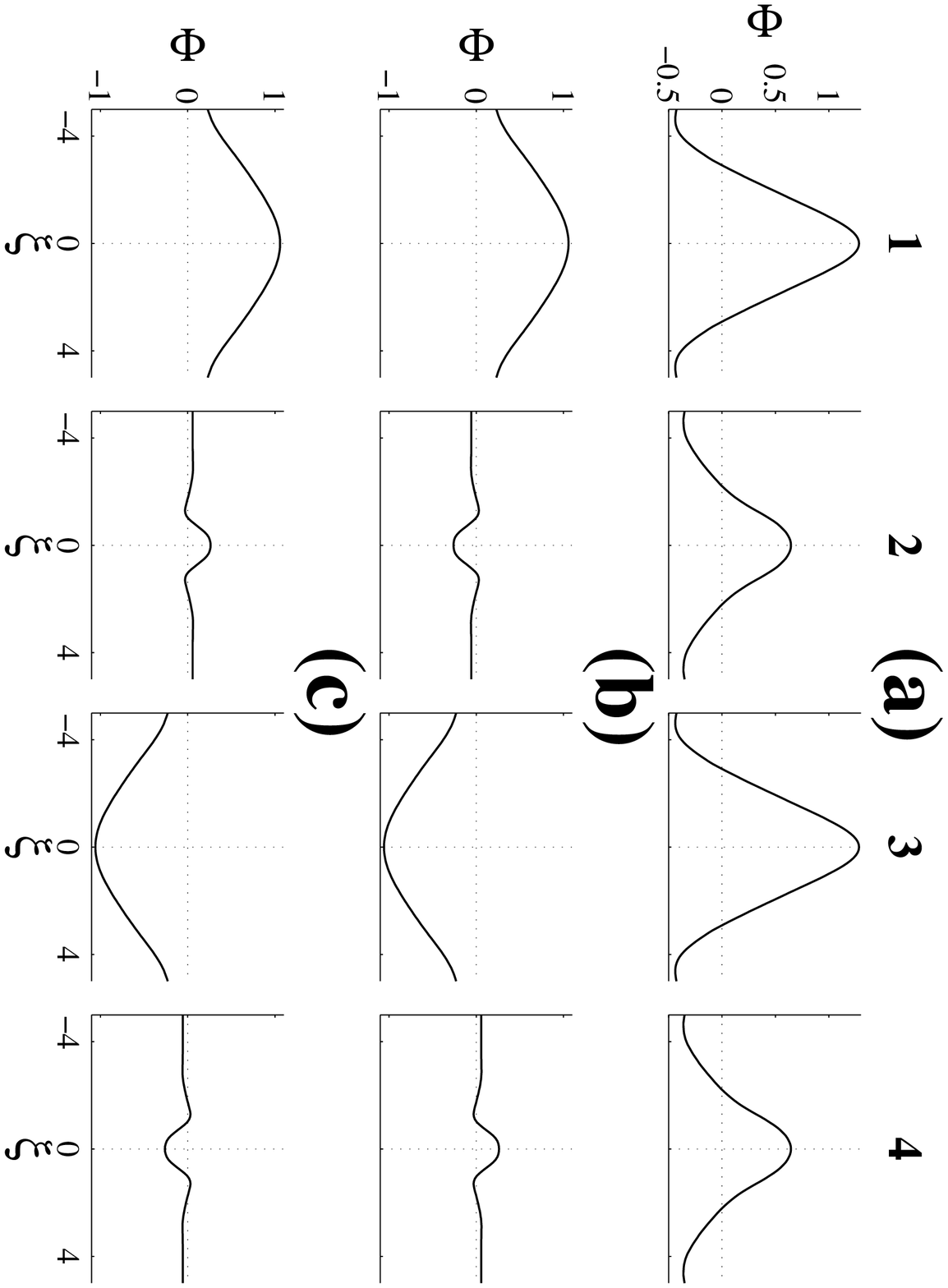,height=18cm}
}
\vspace{1cm}
\caption{\label{f:4p} Solutions of the Eliashberg equations $\Phi$ as function of the bare quasiparticles energy $\xi_k$ for each of the 4 FS pieces and for $\omega_{n=0}$. 
The 2 pieces numbered 1 and 3 correspond to the corners and the 2 others to the planes. 
Different initial conditions for the gap are considered. 
{\bf (a)} With an $s$-wave initial gap, {\bf (b)} with a $d_{x^2-y^2}$ one, and {\bf (c)} with a $d_{xy}$ one. 
The temperature was very low, i.e. $T=0.02$ and the coupling parameters are given in the text. 
(All energies are in units of $\omega_{ph}$).}
\end{figure}

The angular dependence of the Coulomb pseudopotential $\mu(\theta,\theta')$ was set in order to mimic the behaviour observed by neutron scattering \cite{neutron} and the theory Monthoux et al.\cite{pines}. 
This favours a $d$-wave solution. 
However, this enhancement of the repulsive interaction for $|\theta-\theta'| \approx \pi/2$ is not essential for obtaining a $d$-wave. 
Indeed other calculations in the 8 FS pieces case with a constant Coulomb pseudopotential $\mu_{ij} = \mu_0 \;\; \forall i,j$ exhibit a $d$-wave solution quite similar to the one presented in fig. \ref{f:8p}. 
From another calculation (leading to an $s$-wave solution), it seems that the condition for having a $d$-wave solution is that $\lambda_{ij} \leq 2\mu_{ij}$ when $|\theta_i-\theta_j| \approx \pi/2$.

The effect of the repulsive potential $\mu$ on the symmetry of the solution can be seen clearly by comparing fig. \ref{f:4p}a and \ref{f:4p}b (or the two solutions on fig. \ref{f:8p}). 
It can be noted that under the action of the repulsion, the gap parameter $\Phi$ tends to have nodes, i.e. to have different signs at different places of $\mbox{\boldmath $\kappa$}$-space. 
In the $d$-wave case (fig. \ref{f:4p}b), the nodes lie on the FS, located at $\theta_i \approx \frac{\pi}{4} (2n+1)$, and the sign of the gap is mainly constant along $\xi_k$ (for the FS pieces very near the nodes, there are other nodes removed from the FS due to fluctuations and the small value of the gap on these pieces). 
On the contrary, in the $s$-wave case (fig. \ref{f:4p}a), the sign of the gap does not depend on $\theta$ and the nodes are removed from the FS, the change of sign of the gap happening along the $\xi_k$ direction.

In what concerns the nodes for the $d$-wave solution, we want to emphasize that they are not exactly located at  $\theta=\pi/4$ (see fig. \ref{f:8p} and \ref{f:4p}). 
This is due to the relatively small number of FS pieces considered in our calculations. 
But it is nevertheless interesting to notice that this reproduces quite well the situation observed by Ding et al \cite{campuzano} in BiSCCO. 
Indeed they observe the nodes at $\theta = \frac{\pi}{4} \pm \delta\varphi$ and this is compatible with a situation where the experimental result is the average of two degenerate $d$-wave states, one with the nodes shifted by $\delta\varphi$ and the other by $-\delta\varphi$. 
Such a degeneracy may be due to different orientation of the different grains or to orthorhombic distortions. 
However, the $d$-wave symmetry for these materials is still subject to controversy and this explanation does of course not exclude the possibility of an anisotropic $s$-wave or of some effect of the lattice modulation in BiSCCO.

   A salient feature of the present model is, that when the $d$-wave superconductivity
is destroyed, for example by defects or impurities that scatter the conduction 
electrons by a large angle, then $s$-wave superconductivity takes over, without a
large change in the superconducting gap, or in $T_c$. 
Experimentally, we do not 
know yet whether, for example, YBCO with defects is an $s$-wave superconductor; 
but some experiments \cite{klein} favour such a picture. 

   In this picture, the Van-Hove (or extended Van-Hove) singularity (VHS) near the FS
is a secondary factor. 
The $d$-wave pairing exists even for an isotropic FS, 
without any VHS, as long as the scattering angle $\delta\theta$ is sufficiently
small (section \ref{weakcoup}). 
The small $\delta\theta$ also accounts for the high value of
$T_c$ \cite {annphys}. 
In this respect, our model differs from the theory
of Pines and collaborators, and the more recent extension of this theory by
Abrikosov \cite{abrikosov}. 
When a VHS is present near the FS, it favours the 
$d$-wave pairing, causing it to prevail at a somewhat larger value of $\delta\theta$, 
and it further increases the value of $T_c$. 
We believe that the increase in $T_c$
of HgBaCaCuO  from 145 K at ambient pressure, to 164 K under pressure observed
by Chu et al \cite{chu}, is due to the VHS approaching the FS. 
However, the high
value of $T_c = 145$ K at ambient pressure is due predominantly to the small value
of $\delta\theta$.

\section{Experimental Support of the Proposed Model}

   The main assumption of this work is, that the superconducting pairing is 
caused by phonons, and that these phonons have a very small $q$ value, 
$q/2k_F \approx \pi/10$. 
Several experiments lend support to this assumption. 

\begin{enumerate}
\item   A structure is found at an energy of about 40 meV above the gap in YBCO
\cite{vedeneev} and about 45 meV above the gap in BiSCCO \cite{mandrus}.
Vedeneev et al interpret it as a McMillan-Rowell structure due to phonons, and 
extract from it a value of $\lambda$ consistent with the value of $T_c$. 
\item  The value of $\lambda$ extracted from the resistivity of YBCO (and other
cuprates) is very small. 
Martin et al \cite{martin} find $\lambda \approx 0.1$.
The value of $\lambda$ found from band-structure calculations is about 1.7 
\cite{andersen} and the minimum value needed to account 
for $T_c$ of YBCO (assuming a very small value of $\mu^*$) is about 3 \cite{zeyher}. 
The transport $\lambda$ is a factor of 
$\approx \delta \theta^2$ smaller than the McMillan $\lambda$ (this is essentially 
the  $1-\cos\theta$ factor entering the ``classical'' expression for the resistivity).
This indicates that if the superconductivity is indeed due to phonons, 
$\delta\theta$  must be rather small. 
\item A charge density wave (CDW) with wavevector $q=0.24$ $\rm \AA^{-1}$ is observed in BiSCCO overdoped 
with oxygen \cite{hewat}. 
Since $k_F \approx 0.7$ 
$\rm \AA^{-1}$, $q/k_F \approx 0.35$. 
This value of $q$ is close to the value of the 
Thomas-Fermi screening parameter, $\tilde{\kappa}_{TF} = \kappa_{TF}/\sqrt{\epsilon_{ion}}$.
Thus, $\delta\theta \approx \pi/10$. 
This is close to the value of the 
crossover from $d$-wave to $s$-wave pairing described in section \ref{weakcoup}.
Obviously, the estimates made in the present work are rather crude and not quantitative.
Nevertheless, the small value of $q$ provides strong support for the present
model. 
\end{enumerate}

   We wish to point out, that while the observation of a CDW is a proof that 
the electron-phonon interaction is very strong, the converse is not necessarily
true; a very strong electron-phonon interaction causes an appreciable 
softening of the phonons, but not necessarily to $\omega=0$, required for a 
static distortion. 
Therefore, the absence of the CDW in optimally-doped BiSCCO
does not indicate that the electron-phonon interaction there, with a 
$q=0.24$ $\rm \AA^{-1}$ phonon, is not very strong. 
   One possible mechanism to account for the difference in behaviour between 
optimally-doped and oxygen-overdoped BiSCCO, is as follows: In optimally-doped
BiSCCO, the strong electron-phonon coupling softens the LA phonon (for values
of $q$ around 0.24 $\rm \AA^{-1}$), but not to zero, thus there is no static distortion.
The inherently soft TO phonons (with frequencies around 10-15 meV) possess a
different symmetry, and are thus not strongly affected. 
In oxygen-overdoped
BiSCCO, the interstitial oxygens in the $\rm (CuO_2)_n$ planes destroy the symmetry, 
and consequently the LA and TO phonon modes are coupled, and one such coupled
mode softens to zero, producing the static distortion (CDW). 
Thus, the high $T_c$,
and pairing with $d$-wave symmetry, caused by the electron-phonon coupling of the
LA phonon with small $q$, do not necessitate a static distortion; but the static
distortion of the overdoped material points to the very
strong electron-phonon coupling at this $q$ value.

\section{Conclusion}

We have shown in this paper that a phonon-mediated mechanism can lead to a solution of the Eliashberg equations with $d$-wave symmetry, contradicting thereby the conventional belief that it can only lead to $s$-wave solutions. 
This belief is based on the fact that the electron-ion interaction is nearly isotropic in normal metals. 
However the situation is quite different in the cuprates since the large polarizability of the oxygen ionic background prevents an effective electronic screening. 
Therefore, the range of the interaction in momentum space is much smaller than in normal metals. 
This is sufficient to allow for $d$-wave solutions in the weak coupling case (the Coulomb pseudopotential $\mu^*$ being constant, it gives no contribution to the integral in eq. (\ref{eq:bcsgap}) for a $d$-wave gap). 
However, the strong coupling case requires an appropriate repulsion, i.e. a total repulsive interaction, for $|\theta-\theta'| \approx \pi/2$ in order to allow the $d$-wave symmetry. 
This extreme sensitivity to the parameters shows that our solutions of the Eliashberg equations are nearly degenerate and that a small change in the coupling parameters can cause a crossover between the two symmetries.

\newpage

\newpage

\newpage

\newpage

\newpage

\newpage

\begin{thebibliography}{99}
%
\bibitem{pines}
P. Monthoux, A.V. Balatsky, D. Pines, 
Phys. Rev. B {\bf 46}, 14803 (1992)
%
\bibitem{squid}
D.A. Wollman, D.J. Van Harlingen, W.C. Lee, D.M. Ginsberg, A.J. Legget, 
Phys. Rev. Lett. {\bf 71}, 2134 (1993); 
A. Mattai, Y. Gim, R.C. Black, A. Amar, F.C. Wellstood,
preprint;
D.A. Brawner and H.R. Ott, 
Phys. Rev. B {\bf 50}, 6530 (1994); 
C.C Tsuei et al, 
Phys. Rev. Lett. {\bf 73}, 593 (1994); 
%
\bibitem{pokrovski}
V.L. Pokrovski\u{\i} and M.S. Ryvkin, 
Soviet. Phys. JETP {\bf 16}, 67 (1963)
%
\bibitem{combescot}
R. Combescot, 
Phys. Rev. Lett. {\bf 67}, 148 (1991)
\bibitem{peter}
M. Peter, J. Ashkenazi and M. Dacorogna,
Helv. Phys. Acta {\bf 50}, 267 (1977)
%
\bibitem{humlicek}
J. Humli\u{c}ek et al, 
Physica C {\bf 206}, 345 (1993)
%
\bibitem{reagor}
D. Reagor, E. Ahrens, S.W. Cheng, A. Migliori and Z. Fisk,
Phys. Rev. Lett. {\bf 62}, 2048 (1989)
%
\bibitem{vedeneev}
S.I. Vedeneev, P. Samuely, S.V. Meshkov, G.M. Eliashberg, A.G.M. Jansen, P. Wyder,
Physica C {\bf 198}, 47 (1992); 
Y. Yagil, N. Hass, G. Desgardin, I. Monot, 
Physica C {\bf 250}, 59 (1995)
\bibitem{hpa}
B. Barbiellini, M. Weger and M. Peter,
Helv. Phys. Acta {\bf 66}, 842 (1993)
\bibitem{zphys}
M. Weger, B. Barbiellini and M. Peter,
Z. Phys. B {\bf 94}, 387 (1994)
\bibitem{annphys}
M. Weger, B. Barbiellini, T. Jarlborg, M. Peter and G. Santi,
Ann. Physik {\bf 4}, 431 (1995)
%
\bibitem{eliashberg}
G.M. Eliashberg, 
Soviet. Phys. JETP {\bf 11}, 696 (1960); 
{\em ibid} {\bf 12}, 1000 (1961)
\bibitem{miami}
G. Santi, T. Jarlborg, M. Peter and M. Weger,
J. Supercond. {\bf 8}, 405 (1995); 
%
\bibitem{ssp}
G. Santi, T. Jarlborg and M. Peter,
Helv. Phys. Acta {\bf 68}, 197 (1995)
%
\bibitem{zba}
M. Peter, G. Santi and M. Weger, 
to be published
%
\bibitem{jarlborg}
T. Jarlborg and M. Peter, 
J. Physique {\bf 46}, 855 (1985)
%
\bibitem{polonica}
M. Weger, 
Acta Physica Polonica A {\bf 87}, 723 (1995)
\bibitem{arpes}
Z.X. Shen et al,
Phys. Rev. Lett. {\bf 70}, 1553 (1993);
D.S. Dessau et al,
Phys. Rev. Lett. {\bf 71}, 2781 (1993)
%
\bibitem{campuzano}
H. Ding et al, 
Phys. Rev. Lett. {\bf 74}, 2784 (1995)
%
\bibitem{neutron}
J. Rossat-Mignod, L.P. Regnault, C. Vettier, P. Bourges, P. Burlet, J. Bossy, J.-Y. Henry, G. Lapertot,
Physica B {\bf 180-181}, 383 (1992)
%
\bibitem{klein}
N. Klein, N. Tellmann, H. Schulz, K. Urban, S.A. Wolf and V.Z. Kresin,
Phys. Rev. Lett. {\bf 71}, 3355 (1993); 
P. Chaudhari and Shawn-Yu Lin, 
Phys. Rev. Lett. {\bf 72}, 1084 (1994)
%
\bibitem{abrikosov}
A.A. Abrikosov, 
Physica C {\bf 244}, 243 (1995)
%
\bibitem{chu}
L. Gao, Y.Y. Xue, F. Chen, Q. Ziong, R.L. Meng, D. Ramirez, C.W. Chu, J.H. Eggert, H.K. Mao, 
Phys. Rev. B {\bf 50}, 4260 (1994)
%
\bibitem{mandrus}
D. Mandrus, J Hartge, C. Kendziora, L. Mihaly and L. Forro,
Europhys. Lett. {\bf22}, 199 (1993); 
Ch. Renner, {\O}. Fischer, A.D. Kent, D.B. Mitzi, A Kapitulnik, 
Physica B {\bf 194-196}, 1689 (1994)
%
\bibitem{martin}
S. Martin, A.T. Fiori, R.M. Fleming, L.F. Schneemeyer, J.C. Vaszczak, 
Phys. Rev. B {\bf 41}, 846 (1991)
%
\bibitem{andersen}
W. Pickett, Rev. Mod. Phys. {\bf 61}, 433 (1989); C.O.Rodrigues et al,
Phys. Rev. B {\bf 42}, 2692 (1990); T.Jarlborg, Physica B {\bf 172}, 245 (1991)
%
\bibitem{zeyher}
R. Zeyher and G. Zwicknagl,  
Z. Phys. B {\bf 78}, 175 (1990); 
P.B. Allen and D. Rainer, 
Nature {\bf 349}, 396 (1991)
%
\bibitem{hewat}
E.A. Hewat, J.J. Capponi, M. Marezio, 
Physica C {\bf 157}, 502 (1989); 
Y. Le Page, W.R. McKinnon, J.M. Tarascon, P. Barboux, 
Phys. Rev. B {\bf 40}, 6810 (1989); 
A. Yamamoto, M. Onoda, E. Takayama-Muromachi, F. Izumi, T. Ishigaki, H. Asano, 
Phys. Rev. B {\bf 42}, 4228 (1990); 
A. Bianconi, 
Physica C {\bf 235-240}, 269 (1994)
%

\end{thebibliography}
\end{document}